\documentclass[a4paper,fleqn,usenatbib]{mnras}

\usepackage{newtxtext,newtxmath}
\usepackage[T1]{fontenc}
\usepackage{ae,aecompl}
\usepackage{graphicx}
\usepackage{amsmath}
\usepackage{amssymb}

\title[An NS-WD Binary Model for Periodic FRBs]
{A Neutron Star-White Dwarf Binary Model for Periodically Active
Fast Radio Burst Sources}

\author[W.-M. Gu et al.]{
Wei-Min Gu,$^{1}$\thanks{E-mail: guwm@xmu.edu.cn}
Tuan Yi,$^{1}$
and Tong Liu$^{1}$
\\
$^{1}$Department of Astronomy, Xiamen University, Xiamen,
Fujian 361005, P. R. China
}

\date{Accepted XXX. Received YYY; in original form ZZZ}

\pubyear{2020}

\begin{document}
\label{firstpage}
\pagerange{\pageref{firstpage}--\pageref{lastpage}}
\maketitle

\begin{abstract}
We propose a compact binary model with an eccentric orbit to explain
periodically active fast radio burst (FRB) sources, where the system
consists of a neutron star (NS) with strong dipolar magnetic fields
and a magnetic white dwarf (WD).
In our model, the WD fills its Roche lobe at periastron,
and mass transfer occurs from the WD to the NS around this point.
The accreted material may be fragmented into a number of parts,
which arrive at the NS at different times.
The fragmented magnetized material may trigger magnetic reconnection
near the NS surface.
The electrons can be accelerated to an ultra-relativistic speed,
and therefore the curvature radiation of the electrons can account for
the burst activity.
In this scenario, the duty cycle of burst activity is related to
the orbital period of the binary.
We show that such a model may work for duty cycles
roughly from ten minutes to two days.
For the recently reported 16.35-day periodicity of FRB 180916.J0158+65,
our model does not naturally explain such a long duty cycle,
since an extremely high eccentricity ($e > 0.95$) is required.
\end{abstract}


\section{Introduction}
\label{sec1}

Fast radio bursts (FRBs) are millisecond-duration radio pulses from
extragalactic sources, the origin of which remains mysterious
\citep[for reviews, see][]{Katz2018,Petroff2019,Cordes2019,Platts2019}.
The first FRB was discovered by \citet{Lorimer2007}.
Recently, the number of detected FRBs has rapidly increased,
thanks to the Canadian Hydrogen Intensity Mapping
Experiment (CHIME).
The first repeating source, FRB 121102, was reported by \citet{Spitler2016},
and is also the first FRB that has been localized and associated
with a host galaxy \citep{Chatterjee2017}.
During the past two years, 19 new repeating FRBs have been discovered
\citep{CHIME2019a,CHIME2019b,Kumar2019,Fonseca2020}.
It remains a controversy whether all FRB sources are repeating ones.
The absence of repeating bursts even after hundreds of hours of follow-up
and the observed diversity in intrinsic properties
(e.g., temporal structure and polarization)
of one-off FRBs could be evidence for multiple populations of FRBs
\citep{Caleb2018}. 
On the other hand, \citet{Ravi2019} suggests that the volumetric
rate of one-off FRBs exceeds the rate of all possible
cataclysmic FRB progenitors and concludes that most FRB sources are
repeating ones.

Most recently, the first example of periodic activity was reported
by \citet{CHIME2020}, where the source FRB 180916.J0158+65
(hereafter FRB~180916) exhibits an activity period of 16.35 days and the
bursts arrive in a 4.0-day phase window.
FRB~180916 was precisely localized and associated with a star-forming
region in a nearby (redshift $z = 0.0337 \pm 0.0002$), nearly face-on,
massive spiral galaxy with a total stellar mass of approximately
$10^{10}$ solar masses \citep{Marcote2020}.
In addition, no simultaneous event or extended X-ray and gamma-ray emission
was detected according to the recent observations of {\sl AGILE} and
{\sl Swift} \citep{Tavani2020},
as well as {\sl Chandra} and {\sl Fermi} \citep{Scholz2020}.
Obviously, the discovery of periodic activity provides an important
clue to reveal the physics of this repeating FRB.
Some models have been proposed to interpret the periodic behaviour,
such as the precession of a magnetized neutron star or a magnetar
\citep{Yang2020,Levin2020,Zanazzi2020,Tong2020},
the precession of a jet produced by an accretion disc around a massive
black hole \citep{Katz2020},
a mild pulsar in a tight O/B-star binary \citep{Lyutikov2020},
a highly magnetized pulsar whose magnetic field is ``combed" by the
strong wind from a companion star \citep{Ioka2020},
and a pulsar traveling through an asteroid belt \citep{Dai2020}.

In our opinion, even though only one FRB source has well-established
periodicity to date, we may expect that there should exist a population
of periodically active FRB sources.
In this paper, we propose a neutron star-white dwarf (NS-WD) binary model
with an eccentric orbit, where the WD fills its Roche lobe
at periastron, to explain the periodic activity
of a potential population of FRB sources.
The remainder of this paper is organized as follows.
The NS-WD binary model is illustrated in Section~\ref{sec2}.
The relation between the orbital period, the eccentricity, and the WD mass
is studied in Section~\ref{sec3}.
Conclusions and discussion are presented in Section~\ref{sec4}.

\section{NS-WD binary model}
\label{sec2}

In our previous work, \citet{Gu2016} proposed a compact binary model
for repeating FRB sources, which consists of a magnetic
WD and an NS with strong dipolar magnetic fields.
The WD fills its Roche lobe, and mass transfer occurs
from the WD to the NS through the inner Lagrange point.
Magnetic reconnection may be triggered by the accreted magnetized material
when it approaches the NS surface, and therefore the electrons can be
accelerated to an ultra-relativistic speed.
In such a scenario, the characteristic frequency and the timescale of an FRB
can be interpreted by the curvature radiation of the electrons moving
along the NS magnetic field lines.
By considering the conservation of angular momentum and
the gravitational radiation, an intermittent Roche-lobe overflow mechanism
was proposed for the repeating behaviour of FRB 121102.

According to such a model, the duration of an FRB $t_{\rm w}$
can be estimated by the ratio of
the NS radius $R_{\rm NS}$ to the Alfv\'{e}n speed
$v_{\rm A}(=B_{\rm NS}/\sqrt{4\pi \bar\rho})$ \citep{Gu2016}:
\begin{equation}
t_{\rm w} = \frac{R_{\rm NS}}{v_{\rm A}}
= 1.1 \left( \frac{R_{\rm NS}}{10^6 {\rm cm}} \right)
\left( \frac{B_{\rm NS}}{10^{11} {\rm G}} \right)^{-1}
\left( \frac{\bar\rho}{10^3 {\rm g~cm^{-3}}} \right)^{\frac{1}{2}}~{\rm ms} \ ,
\label{eq1}
\end{equation}
where $B_{\rm NS}$ is the magnetic flux density of the NS,
and $\bar\rho$ is the averaged mass density of accreted material.
The above equation indicates that strong magnetic fields ($\ga 10^{11}$~G)
is a necessary condition for the NS. In addition, a WD is the only possible
candidate for the companion in the binary, since the averaged mass density of
a non-degenerate star's atmosphere is significantly lower than
the required typical value
($10^3 {\rm g~cm^{-3}}$) in Equation~(\ref{eq1}).
In other words, a binary consisting of an NS and a non-degenerate star
is unlikely to account for millisecond-duration bursts.

Bursts from FRB 180916 and some other repeating FRB sources exhibit
downward-drifting sub-bursts at a few to tens of MHz~ms$^{-1}$
in the CHIME band \citep{CHIME2019b}, and a characteristic drift rate
of $\sim 200$ MHz~ms$^{-1}$ in the $1.1-1.7$ GHz band
in the case of FRB 121102 \citep{Hessels2019}.
According to our model, such a frequency drift may be qualitatively
understood as follows.
As shown in Equation~(1) of \citet{Gu2016}, the characteristic frequency
of the curvature radiation $\nu_{\rm c}$ is proportional to $\gamma^3$,
where $\gamma$ is the Lorentz factor of the relativistic electrons.
Due to the energy release near the NS surface, the Lorentz factor of
electrons may decrease during the radiative processes, which causes
that the characteristic frequency drifts lower at later times in the
total burst envelope.

\begin{figure}
\includegraphics[width=8cm]{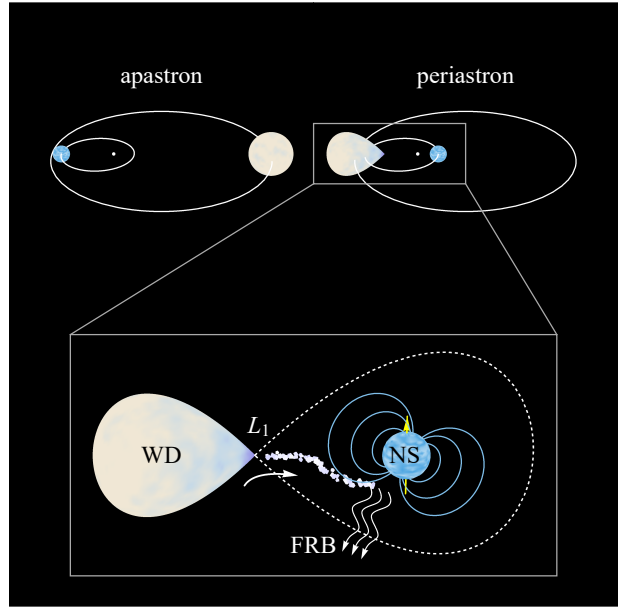}
\caption{
Illustration of the NS-WD binary model with an eccentric orbit,
where the WD fills its Roche lobe at periastron. Around periastron,
mass transfer occurs from the WD to the NS through the $L_1$ point.
The accretion of fragmented material may trigger multiple bursts.
}
\label{f1}
\end{figure}

In the present work, we propose a model following the spirit of \citet{Gu2016}
on the radiative mechanism to explain periodically active FRB sources.
In contrast to
a circular orbit assumption in \citet{Gu2016},
we take the eccentricity into account for an NS-WD binary.
As illustrated in Figure~\ref{f1}, the binary orbit is eccentric
and the WD fills its Roche lobe at periastron. Around this point,
mass transfer occurs from the WD to the NS. The material
of the WD can pass through the inner Lagrange point
($L_1$) and then be accreted by the NS.
Similar to \citet{Gu2016}, such an accretion process can produce FRBs.
For other positions on the eccentric orbit, however, the mass transfer
is interrupted since the Roche lobe is not filled by the WD.
Thus, according to our model, there exists
a window for the bursts in each duty cycle.

We stress that, since the accreted material has angular momentum,
viscous processes are necessary to help the material
lose angular momentum and eventually fall onto the surface of the NS.
During such a process, the accreted material may be fragmented into
a number of parts (as shown in Figure~\ref{f1}), which arrive at the NS
at different times. Thus, a mass transfer
process around periastron may trigger multiple bursts.
In this scenario, the duration of burst activity in each duty cycle is related
to the time interval between the arrival time at the NS of the first part of
the fragmented material and that of the last one,
which can be much longer than the
timescale of the WD passing through periastron.
Thus, the activity window for the bursts can occupy a significant phase
in each duty cycle, as discovered in FRB 180916.

\section{Orbital period}
\label{sec3}

The dynamic equation of the binary (i.e. Kepler's 3rd law)
takes the form:
\begin{equation}
\frac{G(M_{\rm NS} + M_{\rm WD})}{a^3} = \frac{4\pi^2}{P_{\rm orb}^2} \ ,
\label{eq2}
\end{equation}
where $M_{\rm NS}$ and $M_{\rm WD}$ are respectively the NS and WD mass,
$a$ is the binary separation (major axis of the eccentric orbit),
and $P_{\rm orb}$ is the orbital period.
The Roche-lobe radius $R_{\rm L}$ for the WD at periastron can be
expressed as \citep{Eggleton1983}
\begin{equation}
\frac{R_{\rm L}}{a(1-e)} = \frac{0.49 q^{2/3}}{0.6 q^{2/3} + \ln (1+q^{1/3})}
\ ,
\label{eq3}
\end{equation}
where $e$ is the eccentricity of the orbit, and $q$ is the mass
ratio defined as $q \equiv M_{\rm WD}/M_{\rm NS}$.
The WD radius $R_{\rm WD}$ is expressed as \citep{Tout1997}
\begin{equation}
R_{\rm WD} = 0.0115 R_{\sun} \sqrt{(M_{\rm Ch}/M_{\rm WD})^{2/3}
- (M_{\rm WD}/M_{\rm Ch})^{2/3}} \ ,
\label{eq4}
\end{equation}
where $R_{\sun}$ is the solar radius, and $M_{\rm Ch}$ is the 
Chandrasekhar mass limit, $M_{\rm Ch} = 1.44 M_{\sun}$.

\begin{figure}
\includegraphics[width=8cm]{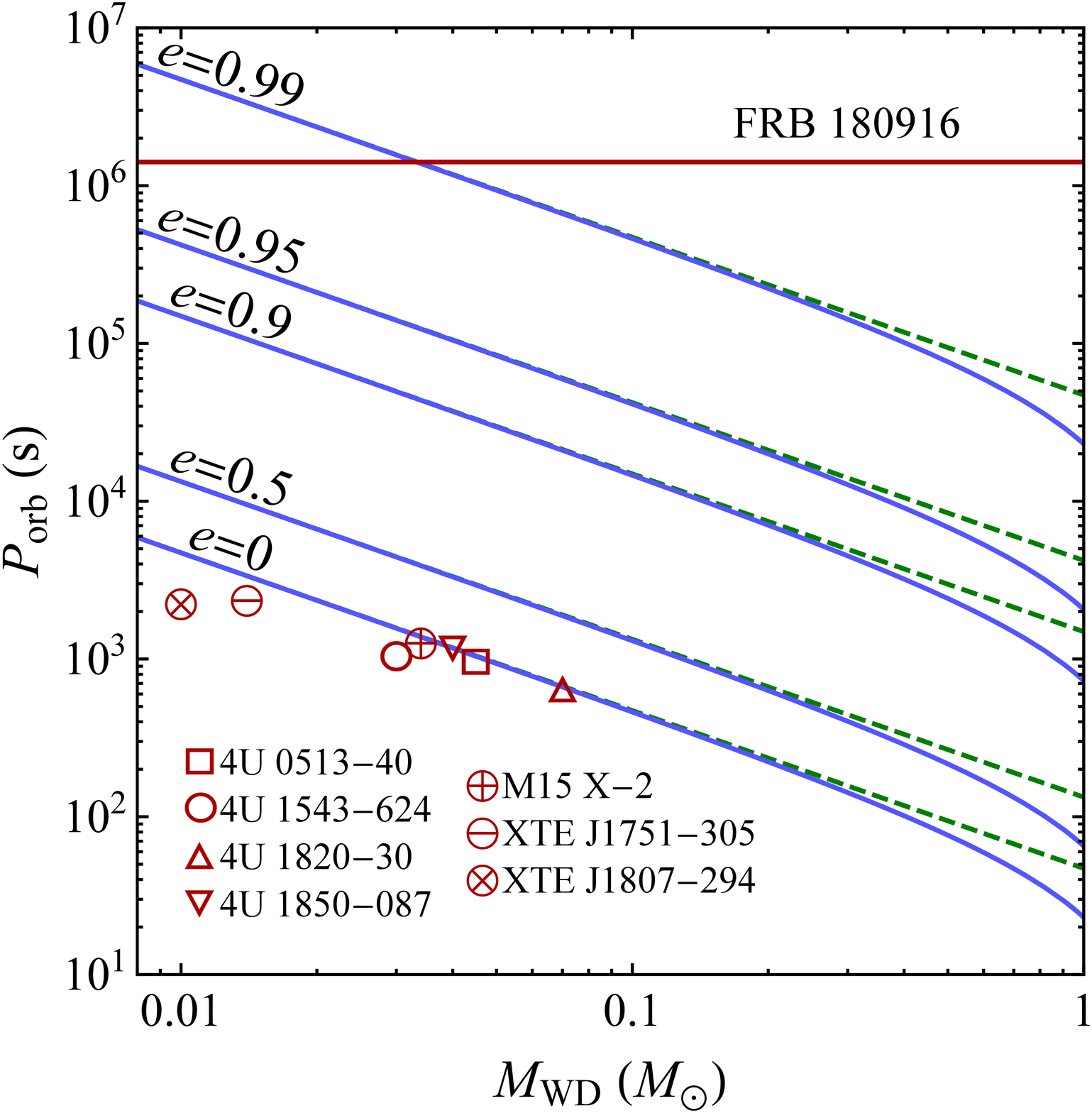}
\caption{
Variation of the orbital period with the WD mass
for five given eccentricities, where $M_{\rm NS} = 1.4 M_{\sun}$
is adopted. The blue curves correspond to
the numerical results calculated using
Equations~(\ref{eq2})$-$(\ref{eq4}),
and the green dashed lines correspond to the analytic relation of
Equation~(\ref{eq7}). The horizontal red line
represents the reported 16.35-day activity period of FRB 180916.
The red symbols denote the seven known UCXBs, which are also contact
NS-WD binaries
and with well constrained orbital periods and WD masses.
}
\label{f2}
\end{figure}

Based on the assumption that the WD fills its Roche lobe at periastron,
i.e. $R_{\rm WD} = R_{\rm L}$, we can derive the values of $P_{\rm orb}$
by Equations (\ref{eq2})$-$(\ref{eq4}) once $M_{\rm NS}$, $M_{\rm WD}$
and $e$ are given.
The variation of $P_{\rm orb}$ with $M_{\rm WD}$ for five given eccentricities,
$e=0$, 0.5, 0.9, 0.95 and 0.99, is shown by five blue solid curves
in Figure~\ref{f2}, where a typical mass $M_{\rm NS}=1.4 M_{\sun}$ is adopted.

The contact NS-WD binaries have been used to explain the
ultra-compact X-ray binaries \citep[UCXBs, see][]{Nelemans2010}.
To our knowledge, seven UCXBs of contact NS-WD binaries
have well constrained orbital periods and WD masses:
4U 1543$-$624 \citep{Wang2004}, XTE J1751$-$305 \citep{Gierlinski2005},
4U 1820$-$30 \citep{Guver2010}, XTE J1807$-$294 \citep{Leahy2011},
4U 1850$-$087, 4U 0513$-$40, and M15 X-2 \citep{Prodan2015}.
In Figure~\ref{f2}, the red symbols represent the seven known UCXBs,
which are well located around the $e=0$ curve.
It is reasonable that their orbits are nearly circular
since mass transfer may cause tidal circularization.
In addition, it is seen from Figure~\ref{f2} that the WD masses of
these seven sources are in the range $(\sim 0.01 - 0.1)~M_{\sun}$.
Recent hydrodynamic simulations on the contact NS-WD binaries
\citep{Bobrick2017} showed that only the systems with
$M_{\rm WD} < 0.2 M_{\sun}$ can exist for a long time, whereas
the systems with $M_{\rm WD} > 0.2 M_{\sun}$
experience unstable mass transfer, which leads to tidal disruption of the WD.
Thus, a reasonable mass range for the WD in a stable contact NS-WD
binary may be $0.01 M_{\sun} < M_{\rm WD} < 0.1 M_{\sun}$.

On the other hand, the discovered eccentric NS-WD binaries may be classified
into two types, i.e., the normal pulsar systems and the millisecond pulsar
(MSP) systems.
For the former, two systems, PSR J1141$-$6545 and PSR B2303$+$46,
have been discovered in our Galaxy \citep[e.g.,][]{Kalogera2005,Sravan2014}.
PSR J1141$-$6545 has an eccentricity $e = 0.17$ and a orbital period
$P_{\rm orb} = 4.744$ hours, and PSR B2303$+$46 has $e = 0.658$ and
$P_{\rm orb} = 296.2$ hours. The relatively long orbital periods indicate that
these two systems are detached binaries.
For the MSP systems, four sources have been discovered
\citep[e.g.,][]{Barr2017}: PSR J2234$+$0611, PSR J1950$+$2414, PSR J0955$-$6150,
and PSR J1946$+$3417. The four systems exhibit similar properties:
the eccentricity $e \approx 0.08-0.14$ and the orbital period
$P_{\rm orb} = 22-32$~days. The long orbital periods also imply that
these systems are detached binaries.
Referring to the above discovered eccentric NS-WD binaries,
a reasonable range for the eccentricity may be $0<e<0.9$.

With a pair of reasonable ranges $0.01 M_{\sun} < M_{\rm WD} < 0.1 M_{\sun}$
and $0<e<0.9$, we obtain the range of $P_{\rm orb}$ roughly from ten minutes
to two days, as shown in Figure~\ref{f2}.
In other words, our compact binary model may work for
the duty cycle of burst activity from ten minutes to two days.
We should note that the eccentricity is not with equal probability
within the range $0<e<0.9$. According to the discovered eccentric NS-WD systems,
we may expect that most eccentric NS-WD binaries have relatively low
eccentricities ($e \la 0.1$), and a small fraction
of NS-WD binaries may have moderate or high values ($e > 0.1$).
Thus, if our mechanism can work for a certain potential population
of repeating FRB sources, then the duty cycles of most sources should exist
in the range roughly from ten minutes to two hours
(corresponding to $e \la 0.1$).

The horizontal red solid line in Figure~\ref{f2} represents the reported
16.35-day activity period of FRB 180916,
which is far beyond the above range of $P_{\rm orb}$.
The red solid line in Figure~\ref{f2} indicates that
an extremely high eccentricity ($e > 0.95$) is required
according to our model. Thus, our model does not naturally explain
such a long duty cycle.

In addition, we may derive a simple analytic relation between
$P_{\rm orb}$, $M_{\rm WD}$, and $e$ as follows.
The Roche-lobe radius $R_{\rm L}$ takes the following simple form
\citep{Paczynski1971} instead of Equation~(\ref{eq3}):
\begin{equation}
\frac{R_{\rm L}}{a(1-e)} = 0.462 \left( \frac{M_{\rm WD}}{M_{\rm NS}
+ M_{\rm WD}} \right)^{1/3} \ ,
\label{eq5}
\end{equation}
and Equation~(\ref{eq4}) may be simplified as
\begin{equation}
R_{\rm WD} = 0.0115 R_{\sun} (M_{\rm CH}/M_{\rm WD})^{1/3} \ .
\label{eq6}
\end{equation}
Thus, we derive the following analytic relation:
\begin{equation}
P_{\rm orb} = 471 \left( \frac{M_{\rm WD}}{0.1 M_{\sun}} \right)^{-1}
(1-e)^{-\frac{3}{2}} \ {\rm s} \ .
\label{eq7}
\end{equation}
Such a simple relation is independent of $M_{\rm NS}$.
The analytic relation is plotted in Figure~\ref{f2} by five green dashed lines.
It is seen that, for the relatively low WD mass region
($0.01 M_{\sun} < M_{\rm WD} < 0.1 M_{\sun}$),
the analytic relation is in good agreement with the numerical results
calculated using Equations~(\ref{eq2})$-$(\ref{eq4}).

\section{Conclusions and discussion}
\label{sec4}

In this paper we have proposed an NS-WD binary model with an eccentric
orbit for periodically active FRB sources.
The WD fills its Roche lobe at periastron and mass transfer occurs
around this point.
The curvature radiation of the ultra-relativistic electrons, which have been
accelerated by the magnetic reconnection, can account for the burst activity.
In this scenario, the duty cycle of the burst activity is related to
the orbital period of the binary.
Based on a pair of reasonable ranges for the eccentricity
and the WD mass, i.e., $0< e < 0.9$ and
$0.01 M_{\sun} < M_{\rm WD} < 0.1 M_{\sun}$,
we have shown that our model may work for the duty cycle roughly
from ten minutes to two days.
For the unique known source with periodic activity, FRB 180916,
our model does not naturally explain such a long duty cycle
since an extremely high eccentricity ($e > 0.95$) is required.

As discussed in \citet{CHIME2020}, one possible explanation for the
discovered 16.35-day periodicity is orbital motion, with either
a stellar or compact-object companion.
The model of a mildly powerful pulsar in a tight O/B-star binary
\citep{Lyutikov2020} hypothesizes that the observed periodicity is due to
absorption of the FRB pulses in the O/B star's wind.
Such a system has a relatively long orbital period ($>$ a few days),
so it is applicable to long duty cycles, such as the
16.35-day periodicity. On the contrary, our model of the NS-WD binary
suggests that the observed periodicity is related to the mass transfer process
around periastron, which is likely to account for relatively short
duty cycles, i.e., less than two days ($e < 0.9$).
We should stress that, if most repeating FRB sources are found to have
long duty cycles of burst activity ($>$ a few days), then our model will
be ruled out.

An open question is on what timescale the orbit can be circularized
to a lower eccentricity due to the accretion around periastron.
\citet{Sepinsky2009} studied the orbital evolution due to mass transfer
in eccentric binaries, by considering the effects of mass and angular
momentum loss from the system.
They found that, when systemic mass and angular momentum loss are
taken into account, the usually adopted assumption of rapid orbital
circularization during the early stages of mass transfer remains unjustified.
According to their results,
the orbital semimajor axis and eccentricity can either increase or decrease,
which is related to the rates of systemic mass and angular momentum loss.
They applied the results to explain the observation of non-zero
orbital eccentricities in mass-transferring binaries, such as a well-known
NS X-ray binary, Circinus X-1, with a 16.6-day orbital period and a high
eccentricity ($e \sim 0.7-0.9$) \citep{Johnston1999} or a moderate
eccentricity ($e \sim 0.4$) \citep{Johnston2016}.
Apart from the accretion process around periastron, gravitational
radiation may also play an important role in the orbital
circularization in close binaries.
The issue of the orbital circularization is beyond the scope
of the present paper, and is worth studying in future works.

\section*{Acknowledgements}

We thank Xiang-Dong Li and Shan-Shan Weng for helpful discussion,
and the referee for constructive suggestion that improved the paper.
This work was supported by the National Natural Science Foundation of China
under grants 11925301 and 11822304.

\bsp
\label{lastpage}

\end{document}